# Terahertz Switch Using an Array of Subwavelength Metallic Holes-coupled-disks

## Sanaz Zarei[1]

School of Electrical and Computer Engineering, College of Engineering, University of Tehran, Tehran, Iran
szarei@sharif.edu

***Abstract*—Broadband switching of terahertz waves at room temperature is demonstrated using a reconfigurable subwavelength metallic hole coupled disk array. The interaction between a metallic membrane featuring periodically arranged circular holes and a substrate bearing a correspondingly periodic array of metallic disks—precisely aligned at their centers—significantly enhances the light coupling within each individual metallic structure, leading to an improved optical transmission and the appearance of a new transmission peak. By mechanical displacement of the metallic membrane with respect to the substrate with metallic disks, the light transmission through the structure can be reconfigured. The device exhibits a polarization-insensitive high-contrast switching performance of 89.4dB at 942GHz. The full-width at half-maximum bandwidth of the switch is 288GHz. By proper design of the device's geometric dimensions, the operation frequency and bandwidth of the switch can be scaled.**

## 1. Introduction

Terahertz research has become highlighted among scientific interests, owing to its broad applicability across diverse domains, including astronomy, biomedical imaging, wireless communications, and defense systems. Despite notable advancements in terahertz systems—particularly in the development of high-performance sources and detectors—the realization of efficient optical components such as waveplates, switches, and modulators for this frequency range is still subject to diligent investigations.

Among these components, terahertz switches are particularly critical for enabling real-time manipulation of terahertz waves, with applications ranging from encoding information in next-generation communication systems to dynamic beam steering and switching for coded aperture terahertz imaging and spectroscopy. Key performance metrics for evaluating a terahertz switch functionality include high switching contrast, minimal insertion loss, and a designable resonance frequency and bandwidth [1].

The extraordinary optical transmission (EOT) phenomenon observed in subwavelength metallic apertures, attributed to the resonant excitation of surface plasmons, has attracted considerable interest for its capability to manipulate electromagnetic waves below the wavelength scale [1]. Surface plasmons have already been demonstrated to play a significant role in integrated photonics, plasmonic devices, nonlinear optics, and biosensing applications [2]. Their scalable nature allows them to operate at almost any desired frequency [2]. The reasonable design of the subwavelength metallic apertures can lead to enhanced localized resonance or high transmission of terahertz waves. Especially multi-layered structures present the possibility of tailoring the resonance properties of the array without reconfiguration of their unit cells, only by changing the arrangements of one layer against the other [1, 3-7].

Several terahertz switches/modulators based on subwavelength metallic apertures incorporating graphene material [8-10], vanadium dioxide film [11], gallium arsenide substrate [2, 12-13], and silicon substrate [14] to demonstrate tunability, have been proposed. Other terahertz switch/modulator schemes include complementary plasmonic arrays [15] and electrically transmission-tunable subwavelength hole array [16]. In this article, dual-layered subwavelength circular arrays in thin metal films are exploited for the demonstration of broadband terahertz switching at room temperature with large isolation and tailored operation frequency.

## 2. Theoretical Background

Surface plasmon polaritons (SPPs) are surface-bound electromagnetic waves that propagate along the interface between a metal and a dielectric [17]. In the visible and infrared spectral ranges, SPPs exhibit strong subwavelength confinement, enabling significant localization of the electromagnetic field at the interface. However, at lower frequencies—such as in the terahertz or microwave regimes—metals behave as near-perfect electric conductors (PECs) because their plasma frequencies lie well above these ranges, typically in the ultraviolet [17]. As a result, conventional SPPs become weakly confined, with fields delocalized over large spatial regions. To overcome this limitation and achieve deep subwavelength confinement at low frequencies, a concept known as "spoof" or "designer" SPPs was introduced. This involves engineering the metal surface—typically through subwavelength-scale structuring, such as periodic grooves or holes—to increase electromagnetic field penetration into the metal and allows tuning the surface plasmon frequency as desired [17]. The existence of spoof SPPs has been experimentally verified in both the microwave and terahertz domains, opening up new possibilities for manipulating light at these scales.

---

[1] Sanaz Zarei is currently affiliated with Research Center for Quantum Engineering and Photonics Technology, Sharif University of Technology, Tehran, Iran

Beyond the bound surface plasmon polariton (SPP) modes, localized surface plasmons (LSPs) are also excited at the periphery of subwavelength apertures [18]. Unlike SPPs, LSP excitation occurs via direct electromagnetic illumination interacting with the curved edges of the aperture [18]. These strongly confined edge-localized modes guide the propagation direction of SPP waves as they tunnel through the aperture, and their coupling with SPPs facilitates enhanced optical transmission [18]. The interplay between these plasmonic modes (SPPs and LSPs) underpins the mechanism of extraordinary optical transmission (EOT), with their dynamic responses at the metal–dielectric interfaces critically influencing the optical properties of EOT light [18]. Modulation of the spectral, spatial, and temporal characteristics of these plasmonic excitations enables coherent control over the EOT effect [18]. The resonance features of both SPP and LSP modes are primarily dictated by the structure geometry and dimensional parameters such as aperture size, spacing, and the dielectric environment [18]. In ultrathin metallic films, SPPs can tunnel through the apertures more efficiently due to interference effects between the top and bottom surfaces [18].

Structured metal surfaces may be divided into two types, one with periodic holes drilled on it (correspondingly inductive grid), and the other with a periodic array of cylindrical metallic pillars (correspondingly capacity grid) [17]. For the surface with periodic holes, the fields of spoof SPPs inside the holes are evanescent waves, whose amplitudes decay rapidly along the hole depth. At the macroscopic scale (i.e., wavelengths greater than the period), the surface with periodic holes acts like a plasma slab, with its plasma frequency matching the cutoff frequency of the holes. Consequently, the confinement of spoof SPPs in such a surface is quite weak and cannot be efficiently improved by increasing the hole depth [17]. In contrast, for the structured surface with periodic pillars, the fields of spoof SPPs in the grid layer are propagating waves as each pair of neighboring pillars in the propagation direction act as a transmission line. Therefore, at the macroscopic level, the grid layer behaves analogously to a dielectric slab and the confinement of spoof SPPs increases with the grid thickness and could become very strong for enough large pillar height [17].

In this article, a terahertz switch consisting of two thin metallic grids of either type is presented. The inductive grid is a thin metallic membrane with a square lattice of circular holes, and the capacitive grid is a square lattice of circular disks supported by a substrate. Each subwavelength metallic hole on the upper membrane is centrally aligned with a subwavelength metallic disk that is located on the substrate. The mutual interaction between the two metallic grids improves the light coupling through each individual grid and leads to an enhanced EOT effect and the emergence of a new transmission peak. Any variation in the spacing between the upper membrane and lower metallic disks results in a change of the coupling strength between the metallic layers and tuning the light transmission through the structure. Previous demonstration of such EOT-enhanced effect through a thin metal film with subwavelength holes blocked by metal disks has been presented in [19]. They reported that in a gold nano-disk coupled nano-hole array, the gold nano-disks collect the incident light like an antenna and induce a significantly enhanced electrical field that is coupled to the nano-hole array and results in an enhanced surface plasmon inside the nano-holes. Larger light transmission up to 70% is observed through the nano-disks coupled nano-holes compared to the bare nano-holes [19].

## 3. Proposed Structure and its Performance

The switch proposed in this work comprises a metal film perforated with a periodic array of circular holes suspended above the substrate [1]. The membrane is integrated with a MEMs actuator, enabling precise longitudinal positioning. Underneath each circular hole, there is a metal disk fixed to the substrate. When the MEMs is actuated by an applied voltage, the suspended metal membrane moves vertically and comes into contact with the disks. Therefore, the transmission of terahertz waves through the structure decreases substantially. The schematic of a unit cell of the switch and its operation principle are depicted in Fig. 1.

The electromagnetic modeling of the switch is carried out using a commercial FEM electromagnetic solver. The structure is modelled as an infinite grid by surrounding its unit cell with periodic boundary conditions. The disks with a radius ($R_d$) of 47.5µm are modeled as a perfect electric conductor (PEC). The metallic membrane perforated with circular holes of radius $R_h$ = 45µm is also modeled as PEC. Both the thickness of the disks ($H_1$) and the membrane perforated with holes ($H_2$) are assumed to be 2µm. The periodicity of the periodic structure (P) is 100µm. As the structure has a four-fold symmetry, its transmission properties are the same for orthogonal polarizations.

The transmission spectrum of terahertz waves through the switch is shown in Fig. 2. When the membrane is located 2µm above the disks (which corresponds to the switch ON-state), a transmission peak at 942GHz with a transmission bandwidth of 288GHz is observed. As the membrane is vertically displaced by the MEMs actuators for an amount of 2µm and thus comes into contact with the disks (switch OFF-state), the transmission peak diminishes. Therefore, by displacing the membrane by an amount of 2µm, the transmission at 942GHz changes more than 85%. The results are obtained for two configurations, one without the supporting substrate and the other with the supporting substrate inserted in the simulations. Polystyrene foam is used as the substrate. The

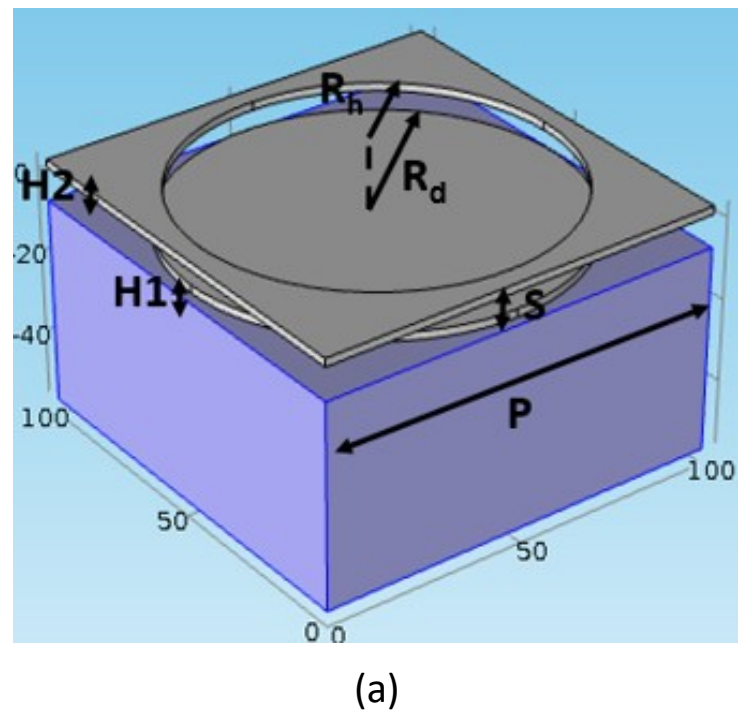

(a)

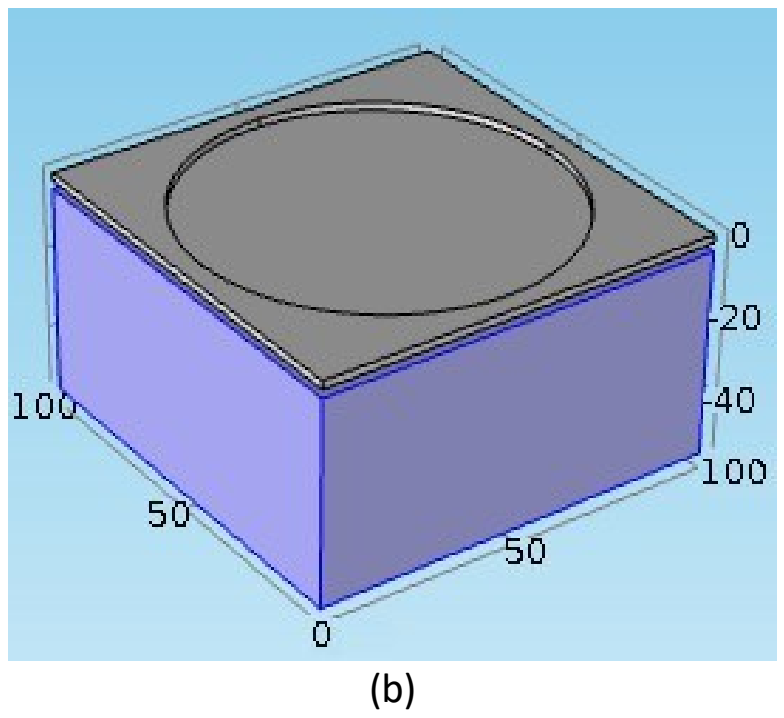
(b)

**Fig. 1** The operation principle of the switch. (a) The membrane is suspended above the disks (ON state). (b) The MEMs is actuated, and the membrane is displaced vertically and comes into contact with the disks (OFF state).

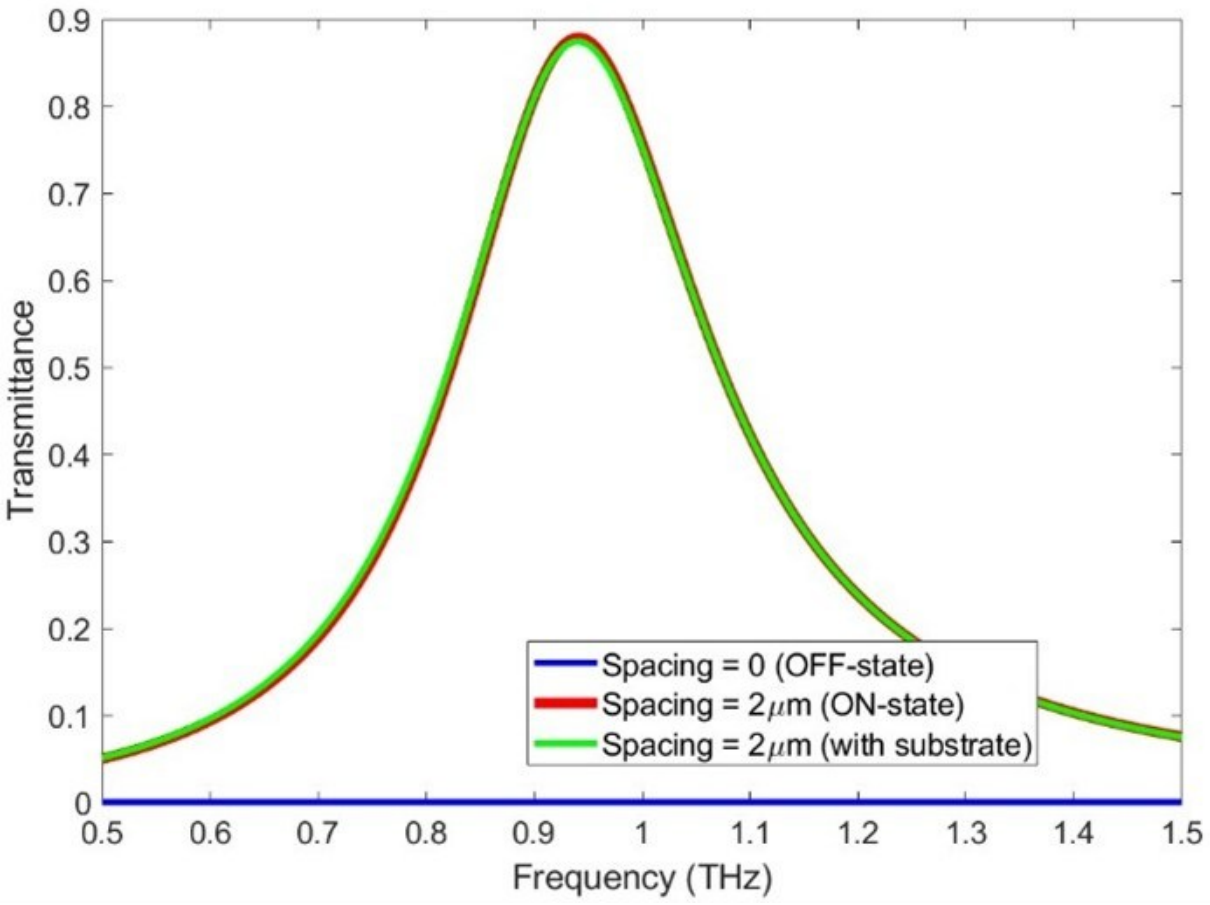

**Fig. 2** (a) The transmission characteristics of the designed terahertz switch for incident terahertz light. The blue curve shows the device transmission at the OFF state, and the red curve shows the device transmission at the ON state. The green curve indicates the transmission at the ON state through the configuration that includes the supporting substrate.

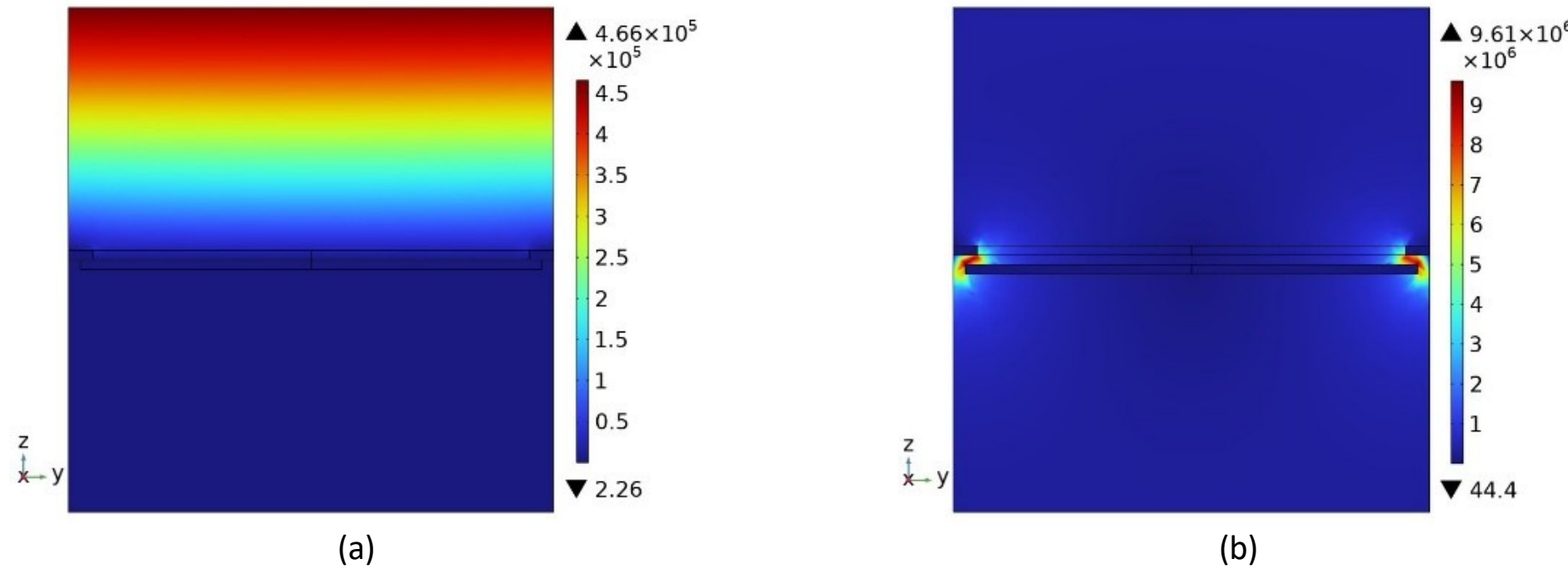

(a) (b)

**Fig. 3** Simulated electric field magnitude of the terahertz switch at 942GHz, corresponding to the peak transmission frequency of the switch at the ON state, (a) when the switch is at the OFF state, (b) when the switch is at the ON state.

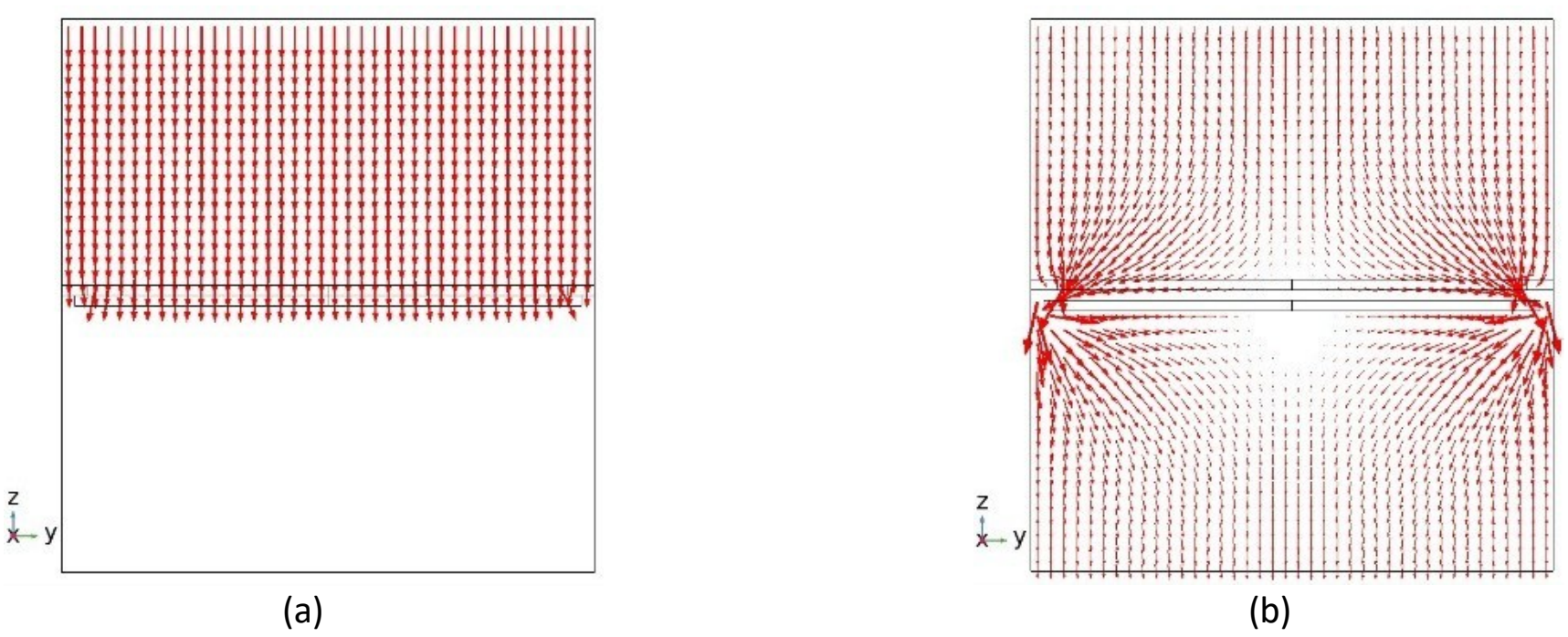
(a) (b)

**Fig. 4** Simulated power flow (time average) through the terahertz switch at 942GHz, corresponding to the peak transmission frequency of the switch at the ON state, when the switch is at (a) the OFF state, (b) the ON state.

proper-choosing of a substrate for the proposed structure is of great importance, as it will dramatically affect the resonance properties of the switch [3-4]. The polystyrene foam is a suitable material for transmitting terahertz radiation because of its remarkably low refractive index of 1.017 to 1.022 and its very low losses in the terahertz frequency range of 0.2-4THz [20]. Based on the four-fold symmetry of the device, it can be realized as a polarization-independent switch through its operation frequency band.

Figures 3(a) and 3(b) plot the simulated electrical field magnitude for the switch OFF-state and switch ON-state at 942GHz, corresponding to the maximum transmission frequency of the hole-coupled disk array. It is clearly seen that the coupling between the metallic hole and the metallic disk significantly enhances the electrical field. This enhanced surface plasmon coupling contributes to the enhanced light transmission to the other side of the metallic structure. The simulated time-averaged power flow within the switch at the OFF-state and On-state is further illustrated in figures 4(a) and 4(b). The bigger arrow sizes and density at the edges of the metallic hole/disk clearly show that the energy flow is highly concentrated near the edges of the hole-coupled disk, which helps to the enhanced transmission through the structure.

At zero bias voltage, which is the switch ON-state, the membrane is suspended above the substrate with a spacing of 2µm. In this state, there is a transmission band around the resonance frequency of 942GHz, for which the insertion loss diagram is depicted in Fig. 5. The insertion loss at the resonance frequency is 0.56dB. When a bias voltage is applied between the substrate with circular disks and the membrane with perforated holes, and the spacing between the two layers becomes zero (switch OFF-state), the transmission band diminishes and the insertion loss reaches the value of 90dB at the resonance frequency of 942GHz, as is shown in Fig. 5.

The switching contrast of the device is 89.4dB at the transmission peak frequency of 942GHz (switch ON-state resonance frequency). The isolation diagram of the switch at other frequencies is shown in Fig. 6. As is seen, the isolation is at least 80dB through the whole transmission band.

## 4. Discussion

As the switch consists of two coupled circular arrays, a metallic membrane perforated with holes, and the metallic disks attached to the substrate, the transmission behavior of the switch at the ON-state can be explained in terms of the light transmission through these two arrays. As is seen in Fig. 7, the disks attached to the substrate act as a low-pass filter, while the metallic membrane perforated with holes acts as a high-pass filter. Therefore, a system composed of both these two arrays is expected to act as a band-pass filter. The maximum transmission of the cascaded system is higher than the anticipated value obtained by the summation of transmission values of the two circular arrays, which is because of the coupling between the two arrays and the enhancement of the electric fields at the sidewalls.

Figure 8 (left side) shows the amplitude transmission of the switch at resonant frequency for different spacings between the metallic membrane and disks. Figure 8 (right side) shows the resonance frequency of the switch and the transmission bandwidth versus the spacing between the layers. It is observed that the transmittance at the resonance frequency of the switch is increased by enlarging the spacing between the layers until it reaches a maximum at 2µm. The same holds for the resonance frequency of the switch. However, the full-width at half-maximum (FWHM) bandwidth of the switch continually increases by the spacing increment. Although it is expected that enlarging the spacing between metallic layers leads to a continuous increment of the transmission amplitude, beyond 2µm, the transmission amplitude drops, which indicates that the coupling between the metallic layers reaches its maximum value at the spacing of 2µm and decreases afterwards.

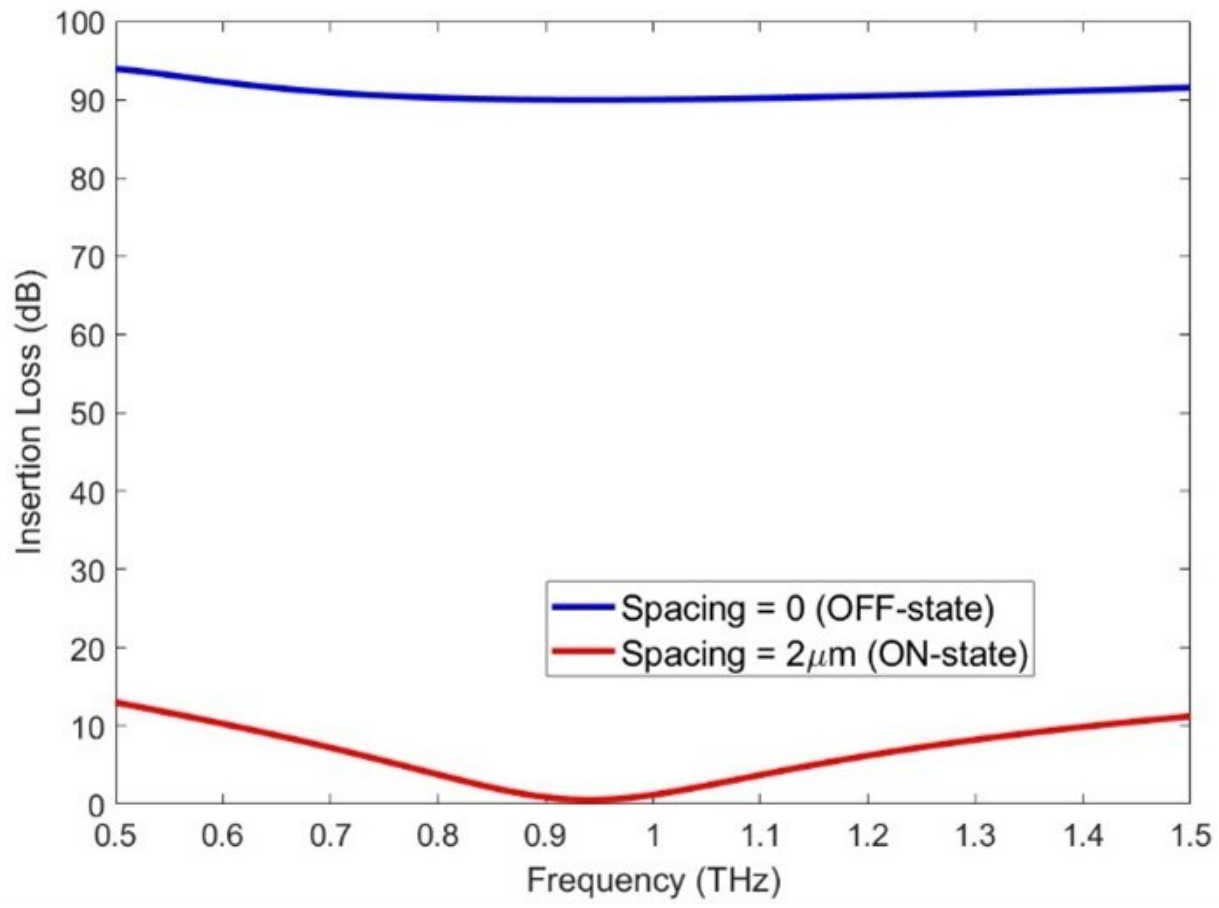

**Fig. 5** The insertion loss of the switch for the ON-state (red curve), and the OFF-state (blue curve).

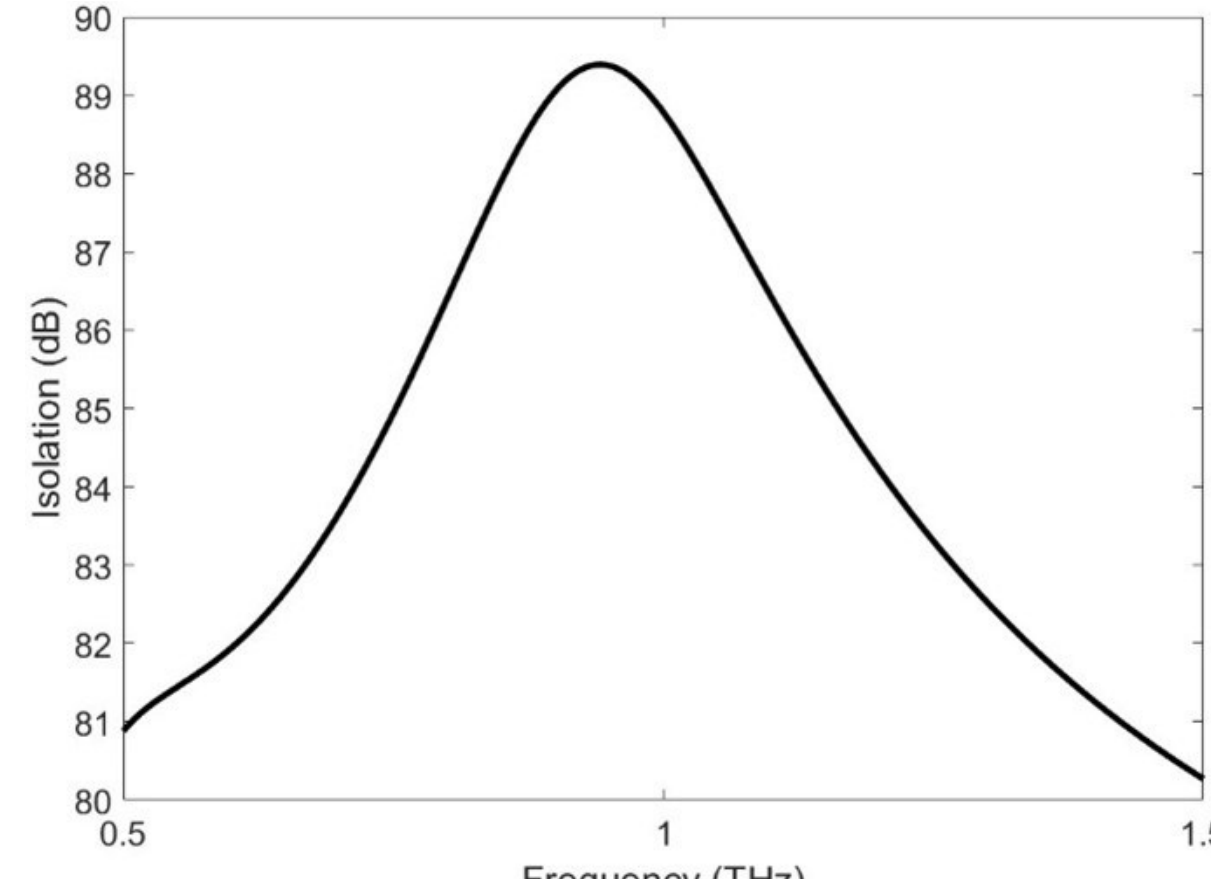

**Fig. 6** The isolation diagram of the switch.

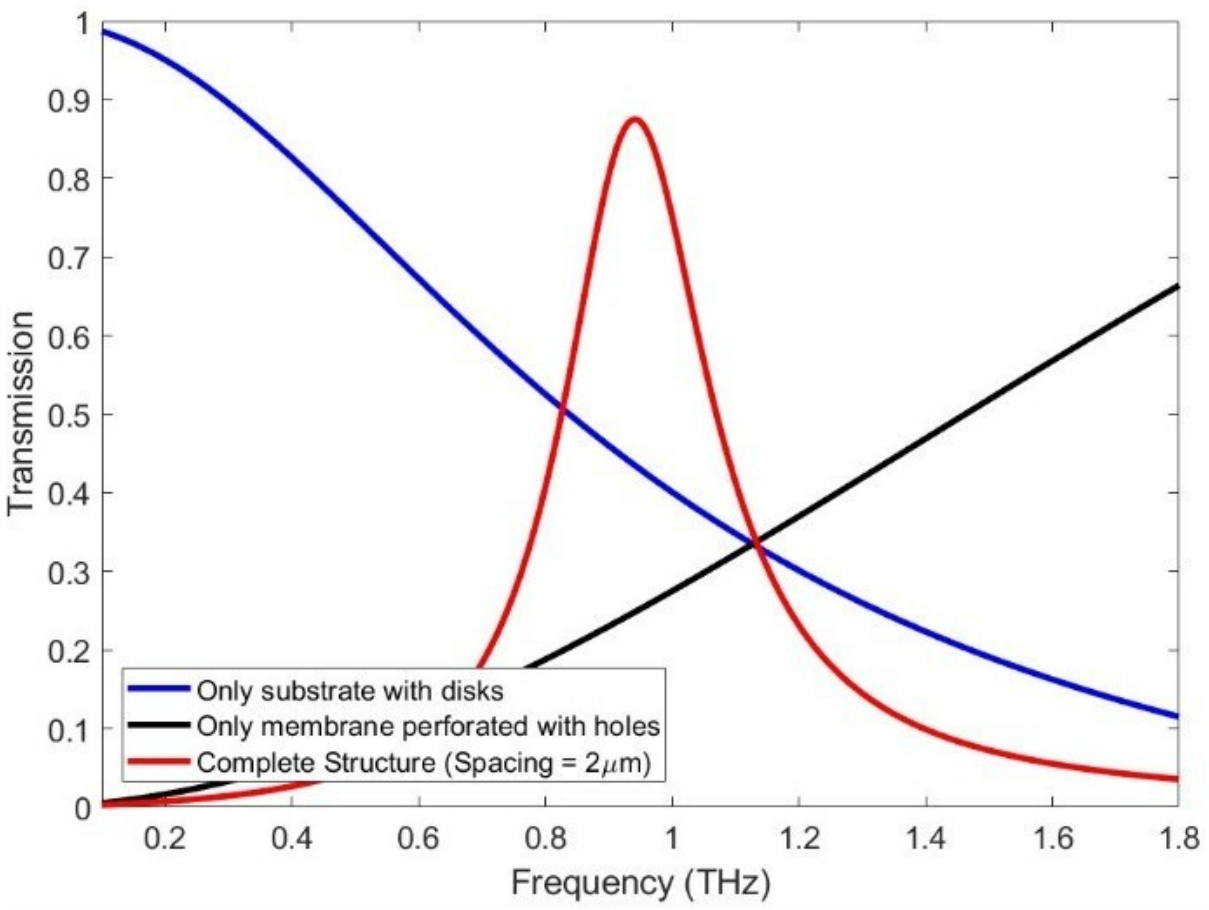

**Fig. 7** The transmission characteristics of the membrane perforated with holes (black curve), the disks attached to the substrate (blue curve), and the presented terahertz switch (red curve) for the incident terahertz light within the frequency band of 0.1-1.8THz. The structural parameters are kept the same as the designed ones for the switch.

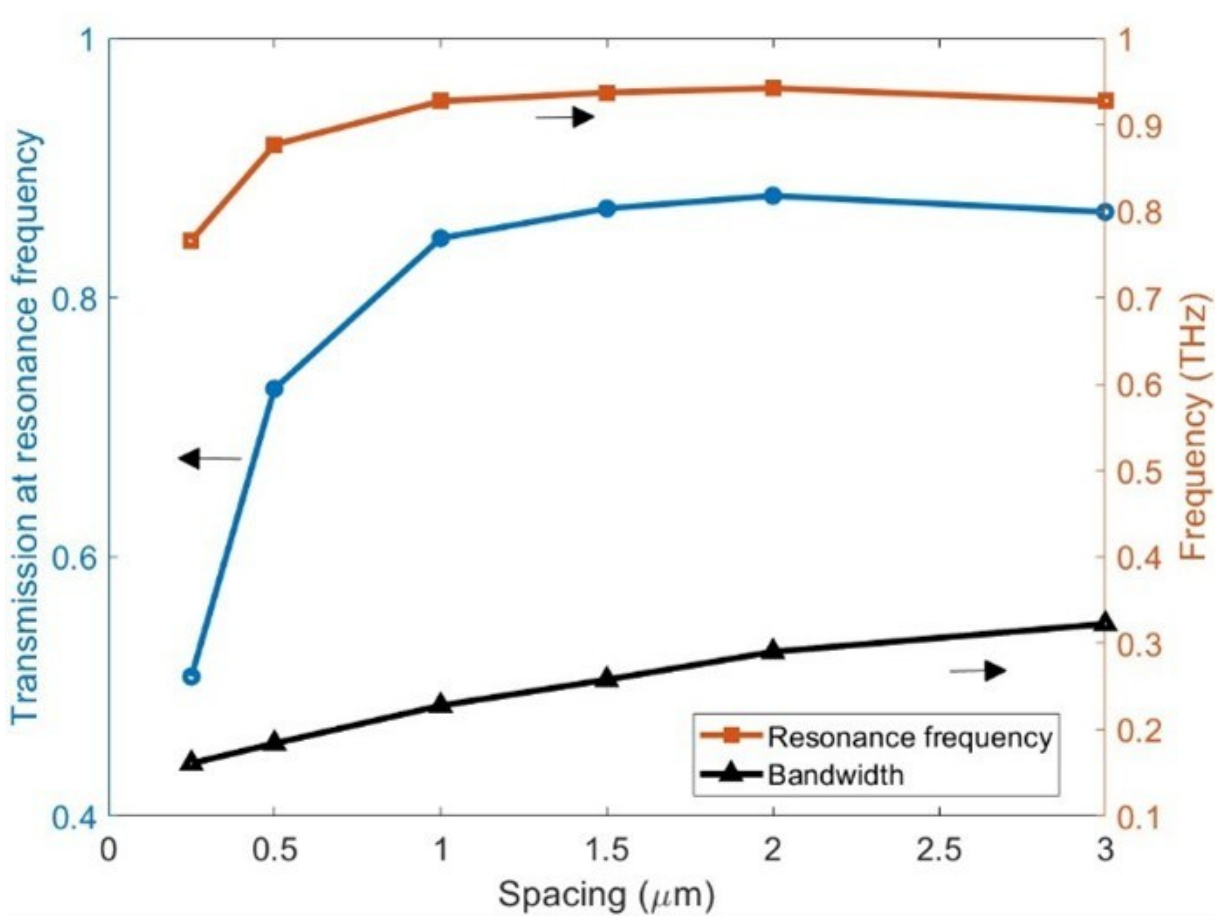

**Fig. 8** (Left side) Amplitude transmission of the terahertz light through the switch at resonance frequency for different spacings between the metallic membrane and disks (blue curve). (Right side) Variation of the resonance frequency (red curve) and FWHM bandwidth (black curve) versus spacing.

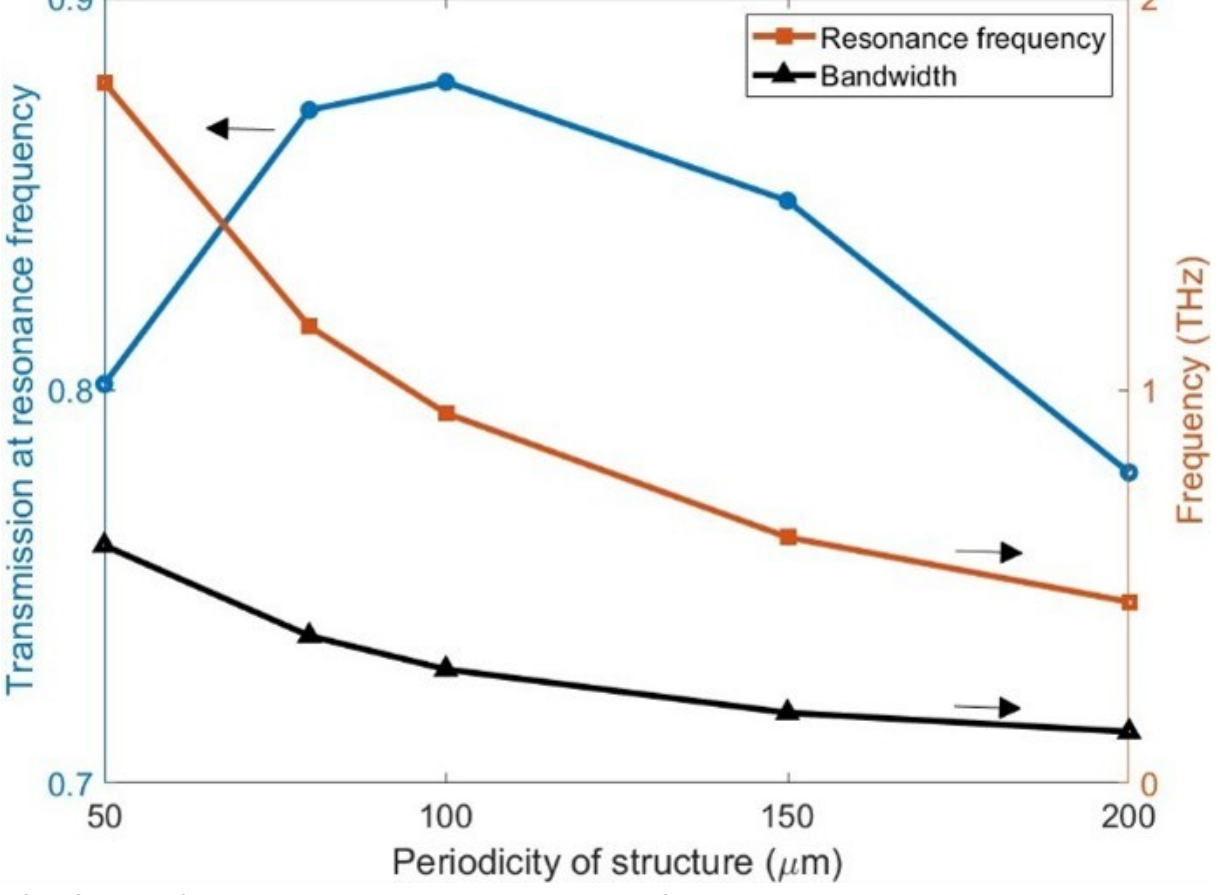

**Fig. 9** (Left side) Amplitude transmission of the terahertz light through the switch at resonance frequency for different values of periodicity of the structure (blue curve), (Right side) Variation of the resonance frequency (red curve) and FWHM bandwidth (black curve) versus periodicity. The height of metallic layers is 2μm, and the ratio of the radius of the holes to that of the disks is kept as 18/19.

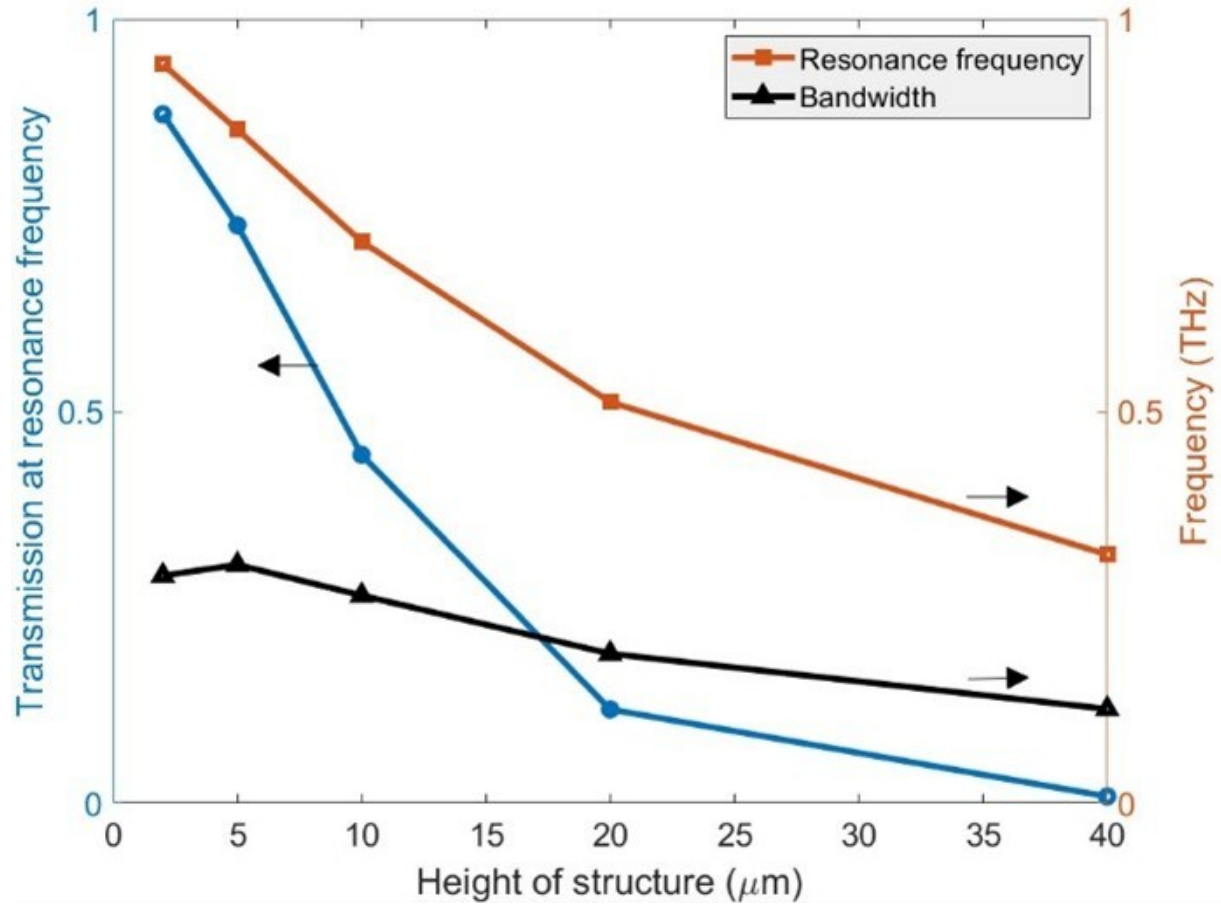

**Fig. 10** (Left side) Amplitude transmission of the terahertz light through the switch at resonance frequency for different values of height of the structure (blue curve), (Right side) Variation of the resonance frequency (red curve) and FWHM bandwidth (black curve) versus height. The period of the structure is 100μm, the radius of the holes is 45μm, and the radius of the disks is 47.5μm.

It can be shown that the switch resonance frequency, bandwidth, and isolation can be tailored by varying the geometry and dimensions of the structure. For the purpose of analyzing the effects of structural parameters on the transmission properties of the switch, the radii of the circles, the structure's height, and the periodicity are changed. As is depicted in Fig. 9 (right side), by increasing the period of the structure, while maintaining the layers' height and radii ratio of the circles, the transmission resonance frequency and bandwidth decrease. Also, from Fig. 9 (left side), the transmission amplitude at the resonance frequency increases until the periodicity reaches 100μm and decreases afterwards. Furthermore, by increasing the height of the structure (i.e., metallic layers) while preserving other parameters, the resonance frequency and bandwidth generally decrease. The transmission bandwidth has its maximum value at the height of 5μm (Fig. 10 (right side)). Also, the transmission amplitude at the resonance frequency decreases with the height increment of the structure, as is depicted in Fig. 10 (left side). In Fig. 10, it is assumed that the height of the metallic membrane and metallic disks are the same.

To further investigate the impact of each circular array's height on the transmission properties, the height of each array is changed individually, as is seen in Fig. 11. The Increment of the metallic membrane's height has a more pronounced role in the variation of the resonance frequency of the switch than the transmission amplitude. However, the increment of the metallic disks' height results in the variation of transmission amplitude more significantly than the resonance frequency.

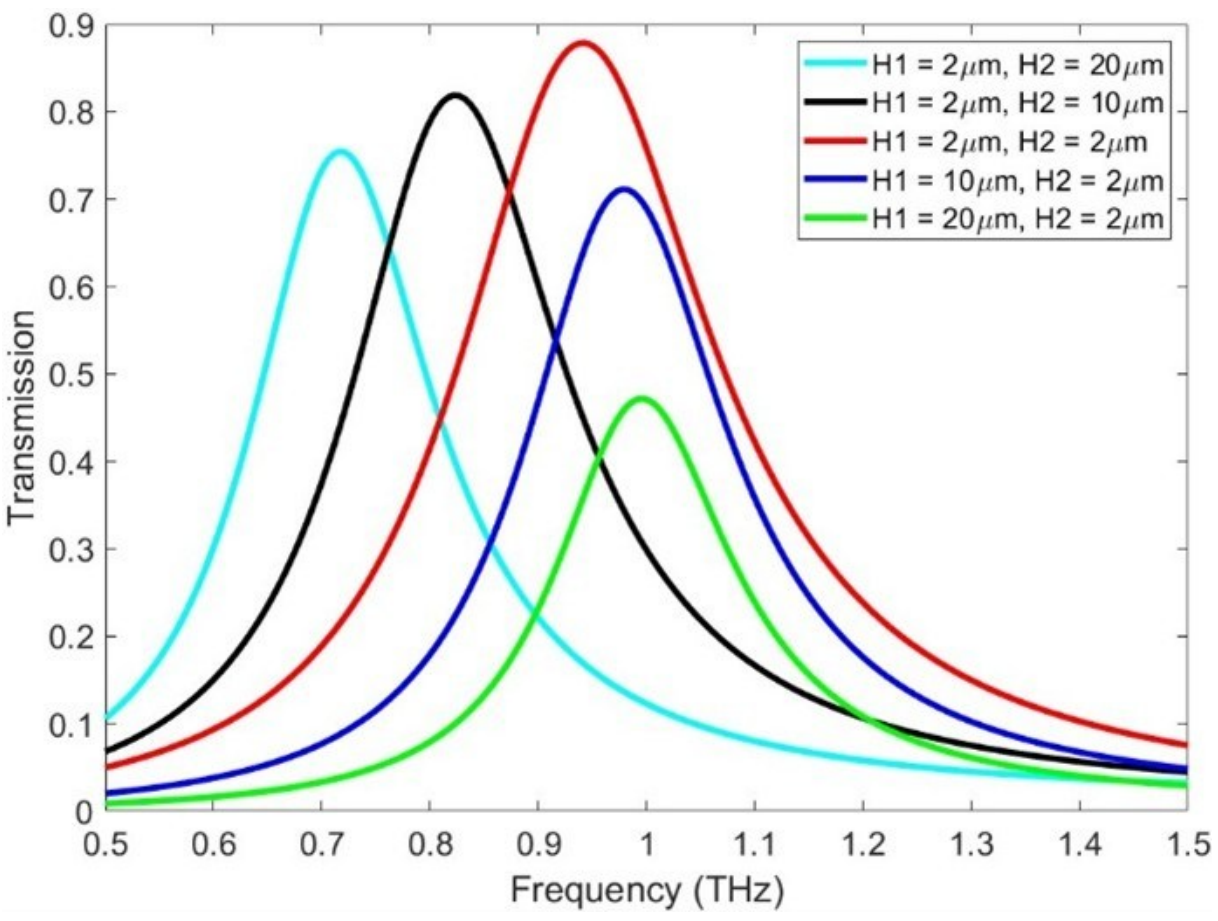

**Fig. 11** The transmission characteristics of the switch for different heights of the metallic membrane ($H_2$) and disks ($H_1$).

Finally, for a fixed radius of the holes, as the disks' radius is increased, the transmission resonance frequency and bandwidth, as well as the amplitude, decrease. On the other hand, for a fixed radius of the disks, by increasing the holes' radius, the transmission amplitude and bandwidth increase, while the transmission frequency decreases.

The present work focuses on the numerical exploration of a terahertz switch using finite-element method (FEM), which directly solve the governing equations and reveal detailed trends that are difficult or impossible to capture analytically. The finite-element numerical framework was validated previously by producing the reported results in [3-5], which were experimentally or analytically validated. Here, to analytically verify the numerical results, the operation of the proposed terahertz switch is interpreted using temporal coupled-mode theory (CMT) complemented by an equivalent circuit model [21]. While full analytical treatment is intractable, a simplified model captures the dominant physics and explains the FEM trends. The array of subwavelength metallic holes coupled to disks supports a dominant hybrid resonant mode that couples to free space through the apertures. Under normal incidence, this system can be modeled as a single resonator coupled symmetrically to two radiation ports. The temporal evolution of the resonant mode amplitude $a(t)$ is governed by:

$$\frac{da}{dt} = (j\omega_0 - \gamma_i - \gamma_e)a + \sqrt{(2\gamma_e)}\, s^+ \tag{1}$$

where $\omega_0$ is the resonance angular frequency, $\gamma_i$ denotes intrinsic losses originating from metallic absorption and substrate dissipation, $\gamma_e$ represents the external radiative coupling through the hole array, and $s^+$ is the incident wave amplitude. The reflected wave amplitude $s^-$ is given by:

$$s^- = s^+ - \sqrt{(2\gamma_e)}\; a \tag{2}$$

Solving the coupled-mode equations in the frequency domain yields the reflection spectrum:

$$R(\omega) = \left|\frac{s^-}{s^+}\right|^2 = \left|1 - \frac{2\gamma_e}{j(\omega - \omega_0) + \gamma_i + \gamma_e}\right|^2 \tag{3}$$

The CMT parameters are directly extracted from the numerically obtained reflection spectra. The resonance frequency $f_0$ is determined from the reflection minimum, while the total decay rate $\gamma_i + \gamma_e$ is obtained from the resonance linewidth. The relative magnitudes of $\gamma_i$ and $\gamma_e$ are evaluated from the minimum reflection value at resonance, enabling independent estimation of both decay rates without curve fitting. The analytical reflection spectra calculated using the extracted parameters show good agreement with the numerical results in terms of resonance frequency, linewidth, and switching contrast, confirming the physical mechanism of the proposed terahertz switch (please see Fig. 12). The simulated reflection spectra at ON-state indicates a single, strongly radiative, low-Q resonance coupled to free space. When the membrane is actuated downward and contacts the disks (OFF-state), the air gap collapses, suppressing the radiative coupling and effectively removing the resonance from the operating band. The CMT model provides a compact physical explanation for the switching behavior.

In parallel, the resonant behavior can be interpreted using an equivalent lumped-element circuit model, which provides intuitive physical insight. In this model, the subwavelength metallic holes act as an effective inductive element due to the confined current flow, while the coupled disks and the intervening gap introduce a capacitive contribution associated with charge accumulation. The hybrid hole–disk resonance can therefore be approximated by an LC circuit with a resonance frequency:

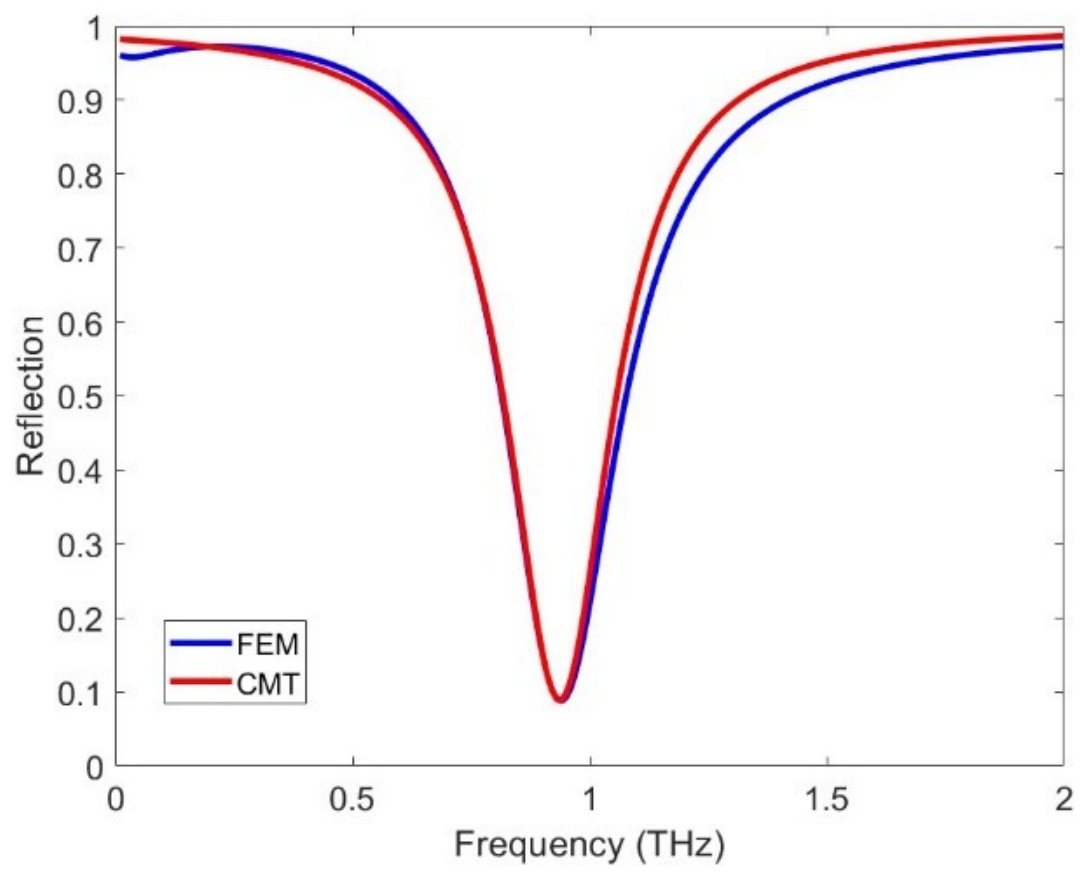

**Fig. 12** The reflection characteristics of the terahertz switch for the incident terahertz light, calculated by FEM (blue curve) and fitted by CMT (red curve).

$$f_0 = \frac{1}{2\pi\sqrt{LC}} \tag{4}$$

In the ON-state (spacing=2μm), the finite capacitance supports a resonance near 942 GHz. When the membrane is actuated downward and contacts the disks (OFF-state), the gap collapses, the capacitance is effectively shorted, and the LC resonance is suppressed, consistent with the observed reduction in transmission. Within this framework, the influence of the geometric parameters can be interpreted. Increasing the disk radius or reducing the disk–hole separation enhances the effective capacitance, leading to a red-shift of the resonance frequency, consistent with the numerical observations. Increasing the hole length or metallic membrane thickness increases the effective inductance, again, causing the resonance frequency to red-shift. Metallic and dielectric losses are represented by an effective resistance in the circuit, corresponding to the intrinsic decay rate $\gamma_i$, while radiation leakage through the apertures is captured by the external coupling rate $\gamma_e$. The switching operation of the proposed device is thus interpreted as a modulation of the effective circuit parameters, primarily the radiative coupling strength. The good agreement between the analytical CMT predictions, the circuit-based interpretation, and the FEM numerical simulations confirms the physical mechanism underlying the terahertz switching behavior.

As a final remark, MEMS actuation is a well-established approach for terahertz devices. MEMS-based reconfigurable components have been widely demonstrated to achieve dynamic control of transmission, phase, polarization, filtering, and switching function in the terahertz range [21-24]. Notably, MEMS-based tunable terahertz filters and modulators have been experimentally demonstrated, and MEMS actuated metamaterials have been realized with measurable modulation depth and switching behavior. Here, MEMS actuation is suggested as one possible implementation, however, the proposed switching mechanism only requires relative displacement between metallic layers and is independent of the specific actuation technology. The required displacement is on the order of 2 μm, which is well within the operational range of existing MEMS actuators. Applicable actuation approaches include, but are not limited to, manual alignment for proof-of-concept demonstrations, electrostatic actuators, thermal micro-actuators, piezoelectric actuators, and wafer bonding with fixed offsets for static ON/OFF states.

## 5. Conclusion

In conclusion, a switch consisting of reconfigurable dual-layered subwavelength circular arrays with a transmission pass-band frequency of 942GHz and bandwidth of 288GHz is proposed. The switching contrast of the device is not less than 80dB and reaches 89.4dB at its resonance frequency. Also, it has a low insertion loss of 0.56dB at its maximum transmission frequency. The switch geometrical parameters have a significant effect on its switching performance, which was extensively discussed in the paper.

## STATEMENTS & DECLARATIONS

### FUNDING

The author declares that no funds, grants, or other support were received during the preparation of this manuscript.

### COMPETING INTERESTS

The author reports there are no competing interests to declare.

### AUTHOR CONTRIBUTIONS

S. Z. performed the study conception and design, accomplished the FEM simulations, analyzed and validated the results, and wrote the manuscript.

## Data Availability

No datasets were generated or analyzed during this study.

## Ethics declarations

This article does not contain any studies involving animals or human participants.